\documentclass{PoS}

\title{Polarisation studies in $H^-t$ production}

\ShortTitle{Polarisation studies in $H^-t$ production}

\author{R. M. Godbole\\
        Center for High Energy Physics, Indian Institute of Science, Bangalore\\
        E-mail: \email{rohini@cts.iisc.ernet.in}}

\author{L. Hartgring\\
        Nikhef, Amsterdam\\
        E-mail: \email{L.Hartgring@nikhef.nl}}

\author{I. Niessen\\
        University of Nijmegen\\
        E-mail: \email{I.Niessen@science.uv.nl}}

\author{\speaker{C. D. White}\\
        \\School of Physics and Astronomy, University of Glasgow\\
        E-mail: \email{Christopher.White@glasgow.ac.uk}}

\abstract{We summarise a recent study looking at top quark polarisation effects in charged Higgs boson production. Lab frame angular and energy observables relating to leptonic decays of top quarks are analysed, and corresponding asymmetry parameters obtained. These are shown to be sensitive to top polarisation and robust with respect to higher order corrections. Thus, they are an efficient probe of charged Higgs parameter space.}

\FullConference{Prospects for Charged Higgs Discovery at Colliders - Charged 2012,\\
		October 8-11, 2012\\
		Uppsala University, Sweden}

\begin{document}

\section{Introduction}
\begin{figure}
\begin{center}
\scalebox{0.8}{\includegraphics{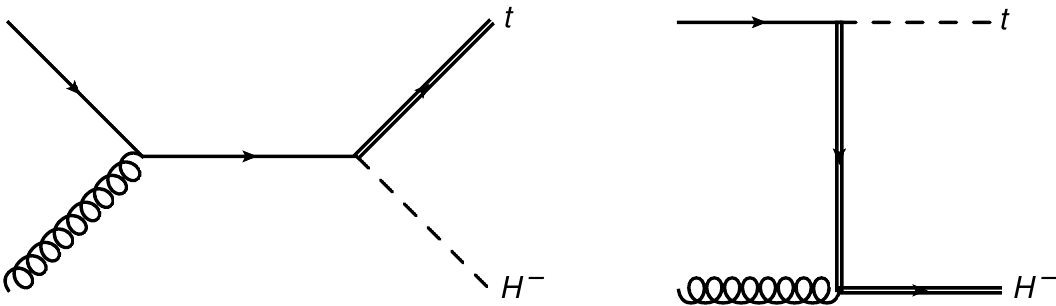}}
\caption{Leading order diagrams for $H^-t$ production.}
\label{fig:LO}
\end{center}
\end{figure}

Many extensions of the Standard Model involve more than Higgs doublet, an immediate consequence of which is the presence of a charged Higgs boson. Observation of the latter would thus be a spectacular confirmation of BSM physics, and the hunt for such particles is ongoing. An important production mode (particularly when the charged Higgs boson mass exceeds that of the top quark) is that of associated production with a (single) top quark. This is analagous to $Wt$ production in the Standard Model, and the leading order Feynman diagrams are shown in figure~\ref{fig:LO}. In this contribution, we assume that the coupling of the Higgs to the top quark is given by a type-II two Higgs doublet model coupling:
\begin{equation}
      G_{H^-t\bar{b}}=-\frac{i}{v\sqrt{2}}V_{tb}\left[m_b\tan\beta(1-\gamma_5)
+m_t\cot\beta(1+\gamma_5)\right],
\label{coupling}
\end{equation}
where, as usual, $\tan\beta$ is the ratio of Higgs VEVs (with overall normalisation $v$); $m_t$ and $m_b$ are the top and bottom quark masses, and $V_{tb}$ the appropriate CKM matrix element. However, our conclusions are more general than this, and may apply in other scenarios.\\

Given that the above coupling involves a superposition of left- and right-handed projectors, the top quark will be produced with a net polarisation on average. The degree of longitudinal polarisation is defined by
\begin{equation}
      P_t=\frac{\sigma(+,+)-\sigma(-,-)}{\sigma(+,+)+\sigma(-,-)},
\label{Ptdef}
\end{equation}
where $\sigma(\pm,\pm)$ is the cross-section for producing a positively or negatively polarised top quark. In SM pair production, each top or antitop is unpolarised on average, so that $P_t=0$. In $H^\pm t$ production, $P_t$ will be nonzero, and will depend on the parameter space $(m_{H^-},\tan\beta)$. If the top
quark decays according to $t\rightarrow Wb\rightarrow f\bar{f}b$ for some fermionic decay product $f$, the latter is distributed in the top quark rest frame
\begin{displaymath}
\sim\frac{1}{2}\left(1+\kappa_fP_t\cos\theta_{f,\rm{rest}}\right).
\end{displaymath}
Here $\theta_{f,\rm rest}$ is the angle between $f$ and the top quark spin vector, and $\kappa_f$ the {\it analysing power}, whose value depends on the identity of $f$, and potentially receives higher order corrections from both SM and BSM sources. For a positively charged lepton ($f=l$), the analysing power $\kappa_l\simeq 1$, and turns out to be insensitive to BSM corrections to the decay of the top quark, up to quadratic corrections in the inverse scale of new physics~\cite{Godbole:2010kr}. Thus, leptonic decay products of top quarks are highly efficient polarisation analysers. \\

Consider a new physics particle $X$ produced in association with a top quark. The coupling of $X$ to the top will produce a non-zero polarisation $P_t\neq0$ in general. If the top decays leptonically, the angular distribution of the produced lepton will be governed by $P_t$ and $\kappa_l$. BSM corrections to the decay are irrelevant to a good approximation, as stated above, and thus it follows that leptonic decay products of top quarks can be used to directly probe the coupling of $X$ to the top quark. Here $X$ is a charged Higgs boson, but our argument is easily generalised to other new physics scenarios.\\

Above we considered the top quark rest frame. Given the difficulty of fully reconstructing top quark momenta, it is easier and more useful to consider quantities in the lab frame. Assuming that the top quark direction can be reconstructed, one may define the azimuthal ($\phi_l$) and polar ($\theta_l$) angles between the decay lepton and the direction of its parent top. These were first defined in~\cite{Godbole:2010kr}, and analysed within a charged Higgs context in~\cite{Huitu:2010ad,Godbole:2011vw} (see also~\cite{Baglio:2011ap}). Both carry an imprint of top polarisation: a purely polar correlation in the lab frame appears as a mixture of polar and azimuthal information in the lab frame, due to the non-zero boost of the top quark in the latter.\\

\section{Results in $H^-t$ production} 
In~\cite{Godbole:2011vw}, we examined the above angular distributions using the recently developed MC@NLO software for $H^-t$ production~\cite{Weydert:2009vr}. This combines NLO matrix elements for this production process with parton shower and hadronisation algorithms, and also includes spin correlations~\cite{Frixione:2007zp}. By comparing these results with LO ones obtained using MadGraph~\cite{Alwall:2011uj}, one may evaluate the robustness of polarisation observables with respect to higher order corrections. All results here were obtained with top and bottom quark masses of $m_t=172$ GeV and $m_b=4.95$ GeV respectively, a top width of $\Gamma_t=1.4$ GeV, and a common renormalisation and factorisation scale $\mu_r=\mu_f=m_t$. We use MSTW2008 LO and NLO partons~\cite{Martin:2009iq}, as appropriate. 

\begin{figure}
\begin{center}
\scalebox{0.35}{\includegraphics{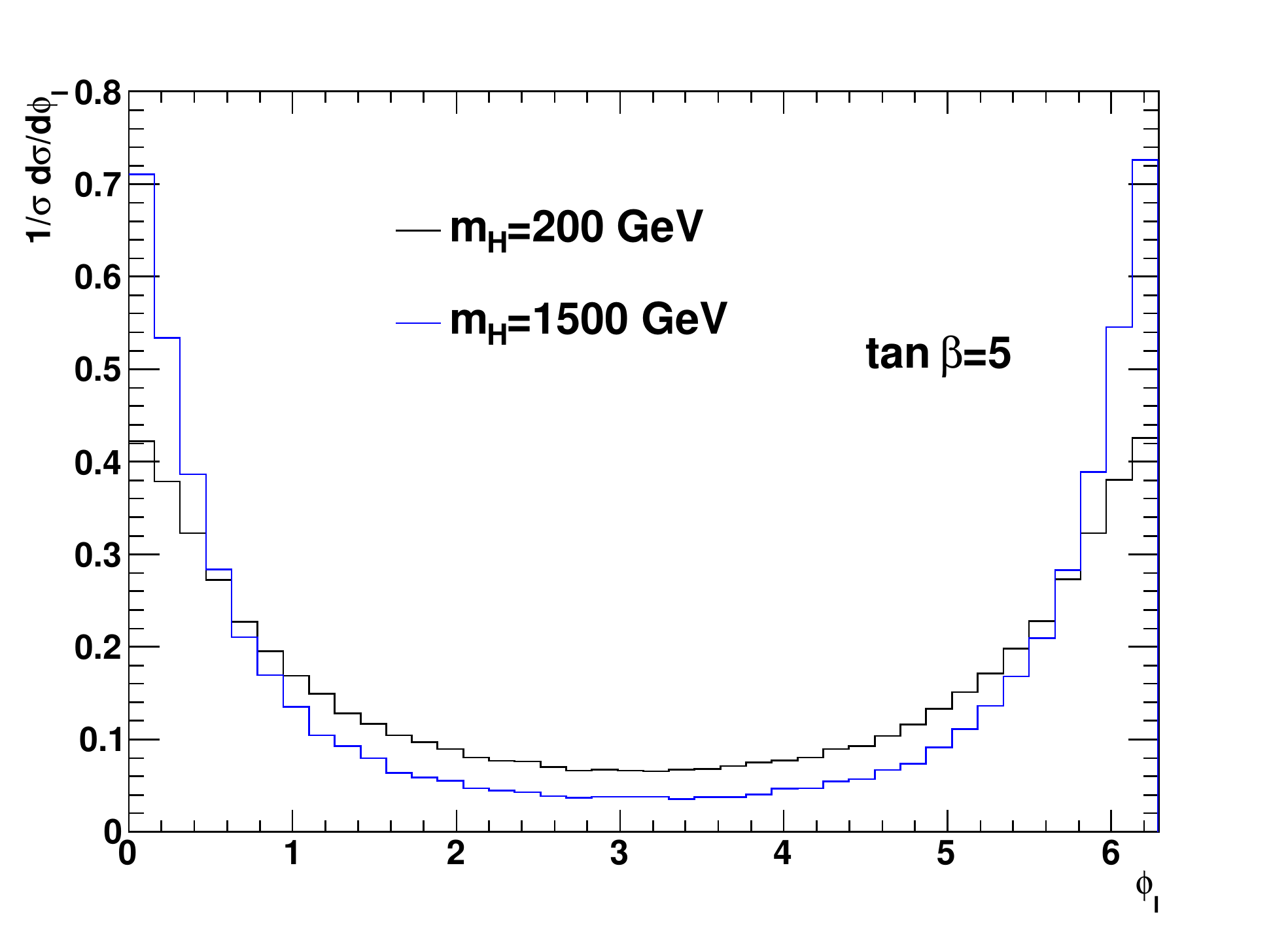}}
\scalebox{0.35}{\includegraphics{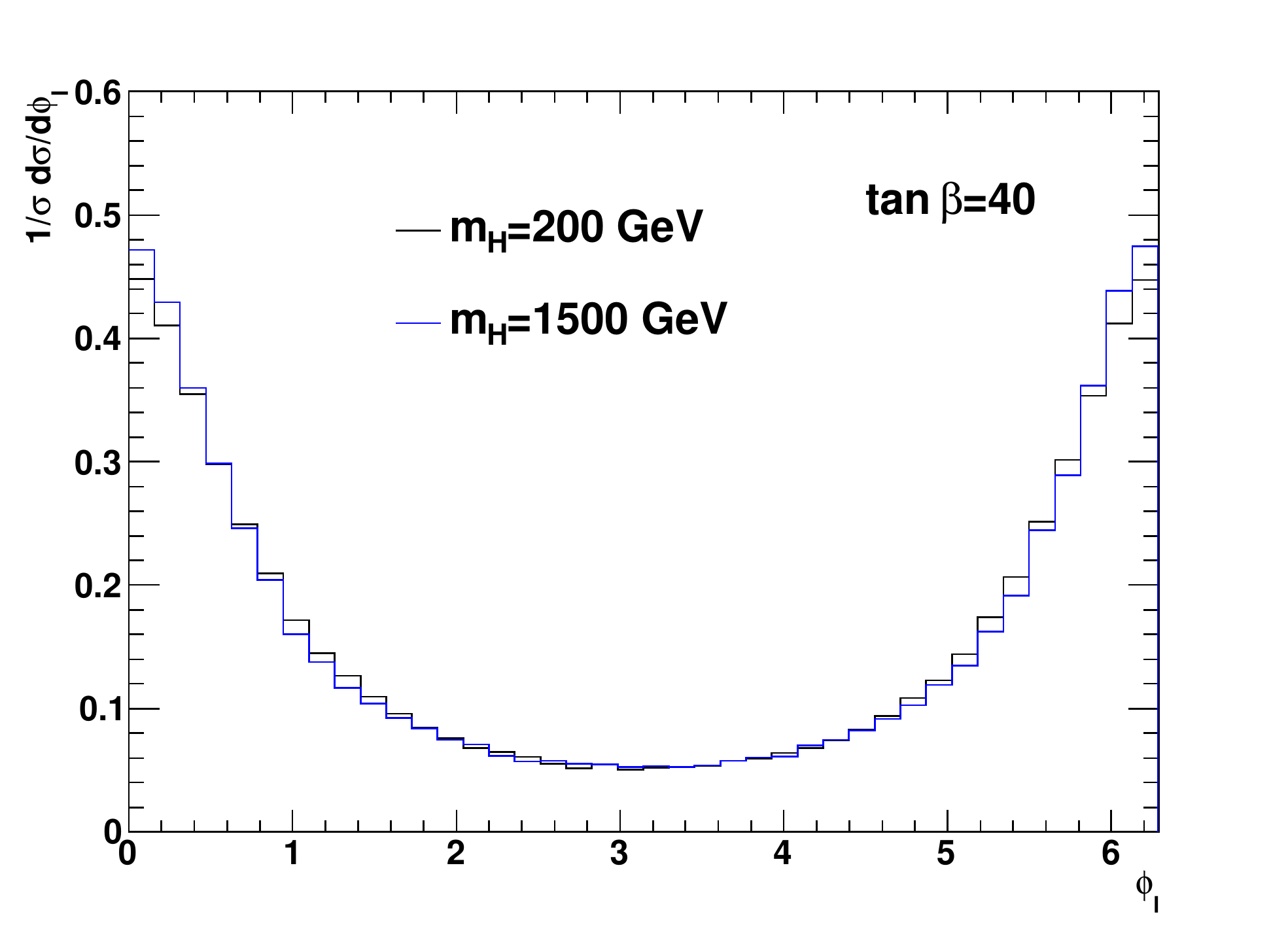}}
\caption{Azimuthal angle $\phi_l$ between the top quark and its decay lepton in $H^-t$ production, obtained using MC@NLO~\cite{Weydert:2009vr}.}
\label{phiplot}
\end{center}
\end{figure}

\begin{figure}
\begin{center}
\scalebox{0.35}{\includegraphics{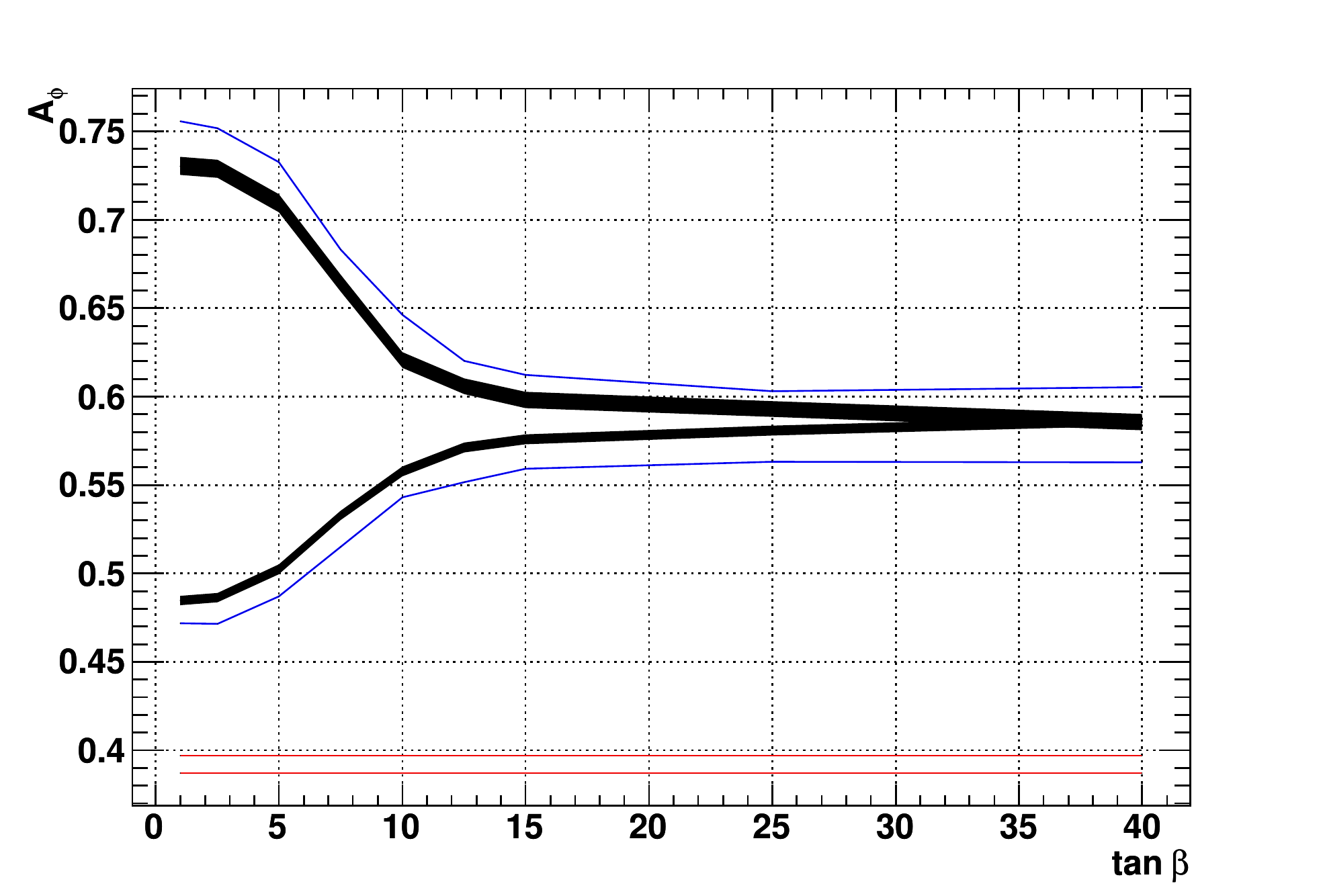}}
\scalebox{0.35}{\includegraphics{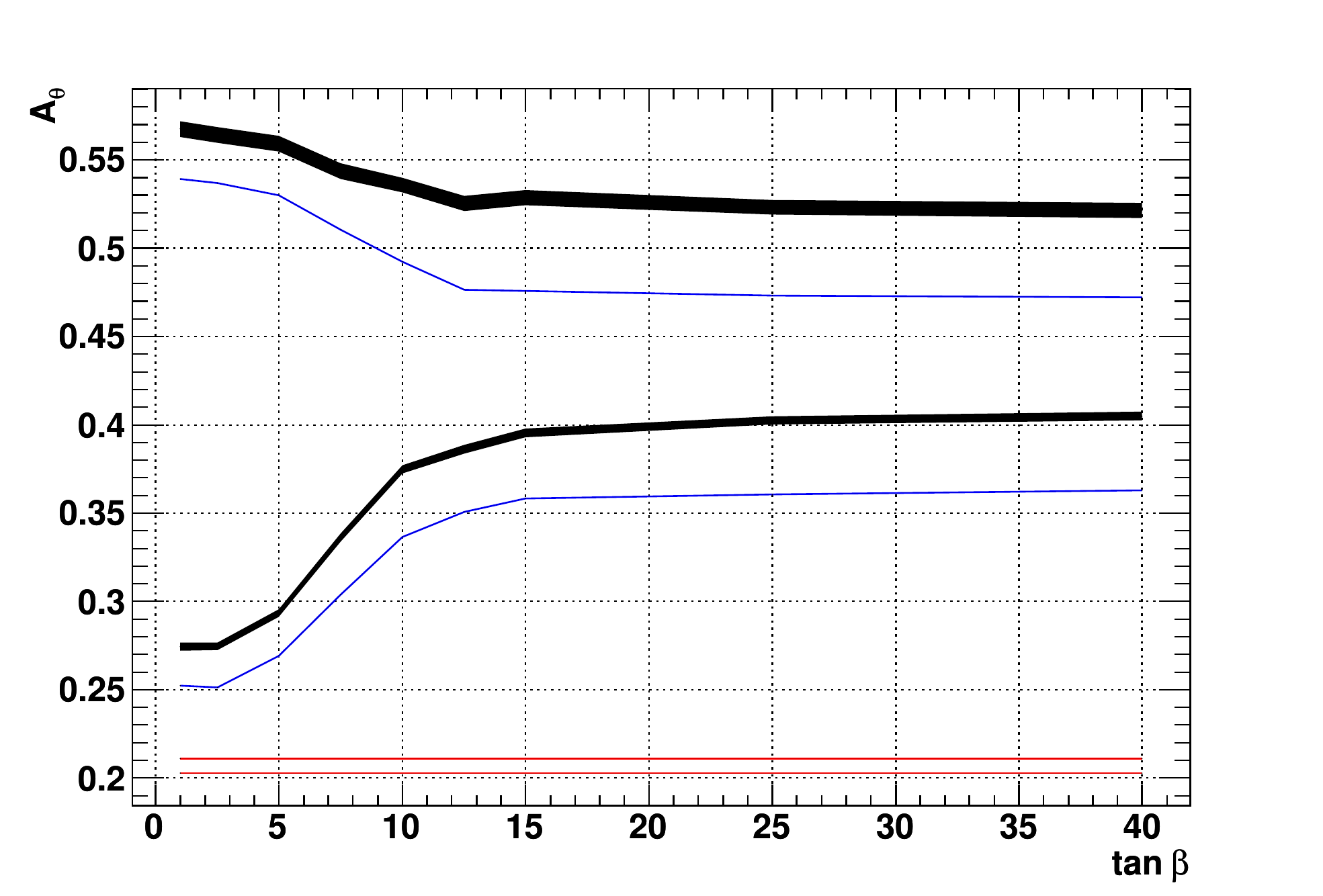}}
\caption{Behaviour of (a) the azimuthal asymmetry parameter $A_\phi$; (b) the polar asymmetry parameter $A_\theta$, at both LO (blue) and MC@NLO (black) levels. Upper and lower curves correspond to charged Higgs boson masses of 1500 and 200 GeV. Also shown are the results obtained for SM $Wt$ production (lower bands).}
\label{aplot}
\end{center}
\end{figure}

In figure~\ref{phiplot} we show the distribution of the azimuthal angle $\phi_l$, as a function of $\tan\beta$ and for two different charged Higgs boson masses. It is strongly peaked at $\phi_l=0$ and $\phi_l=2\pi$, due to the boost from the top quark rest frame. Polarisation information modifies this shape, and we can efficiently distil this information into a single number: noting the crossing points in figure~\ref{phiplot}, we can define the asymmetry parameter
    \begin{equation}
      A_{\phi}=\frac{\sigma(\cos\phi_l>0)-\sigma(\cos\phi_l<0)}{\sigma(\cos\phi_l>0)+\sigma(\cos\phi_l<0)}.
    \end{equation}
Each point in parameter space corresponds to a different value of $A_\phi$, as
shown in figure~\ref{aplot}(a). Thus, this quantity can be used to pin down the coupling of a charged Higgs boson to the top quark, as well as to enhance the signal to background ratio. \\

\begin{figure}
\begin{center}
\scalebox{0.35}{\includegraphics{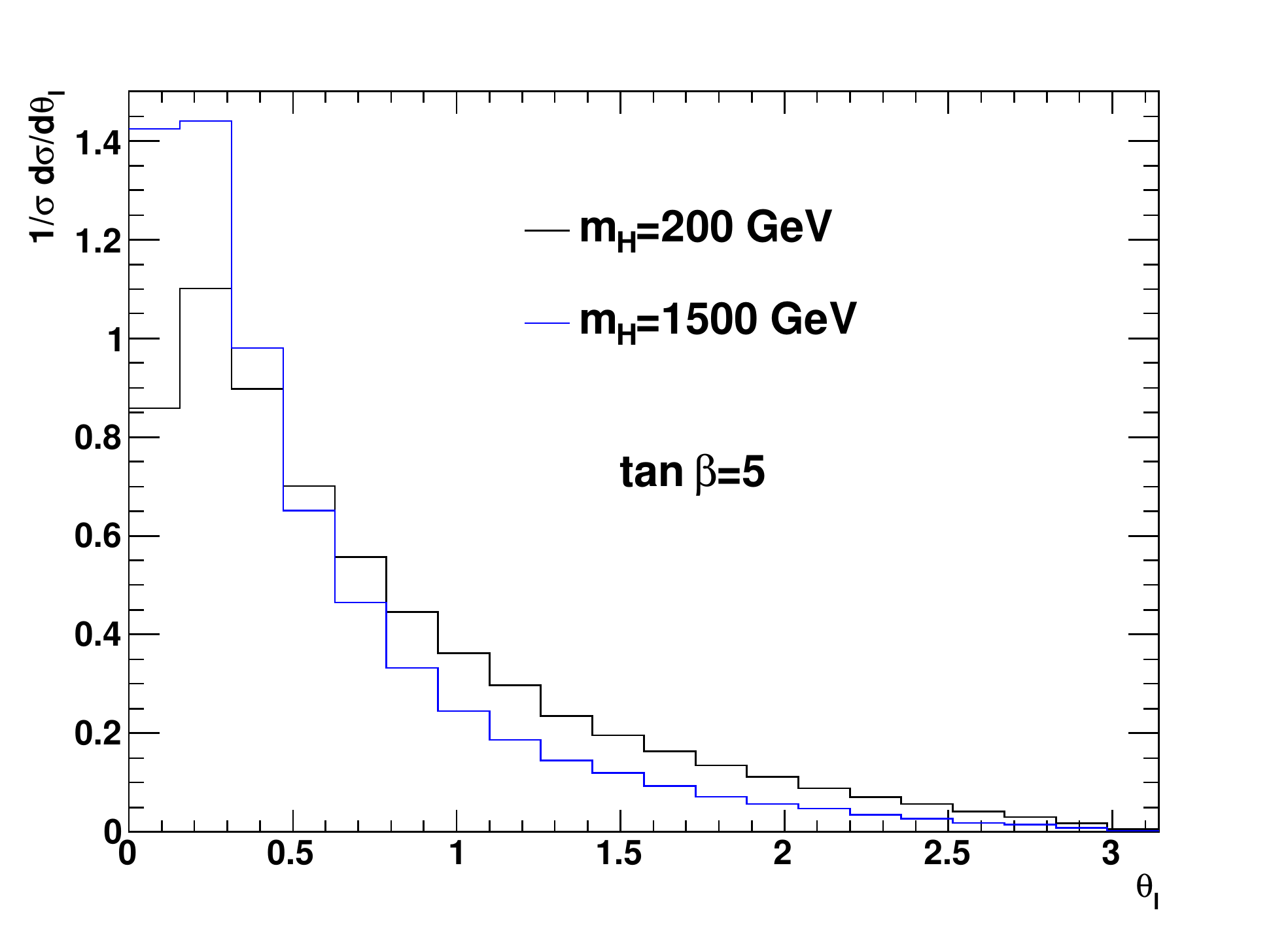}}
\scalebox{0.35}{\includegraphics{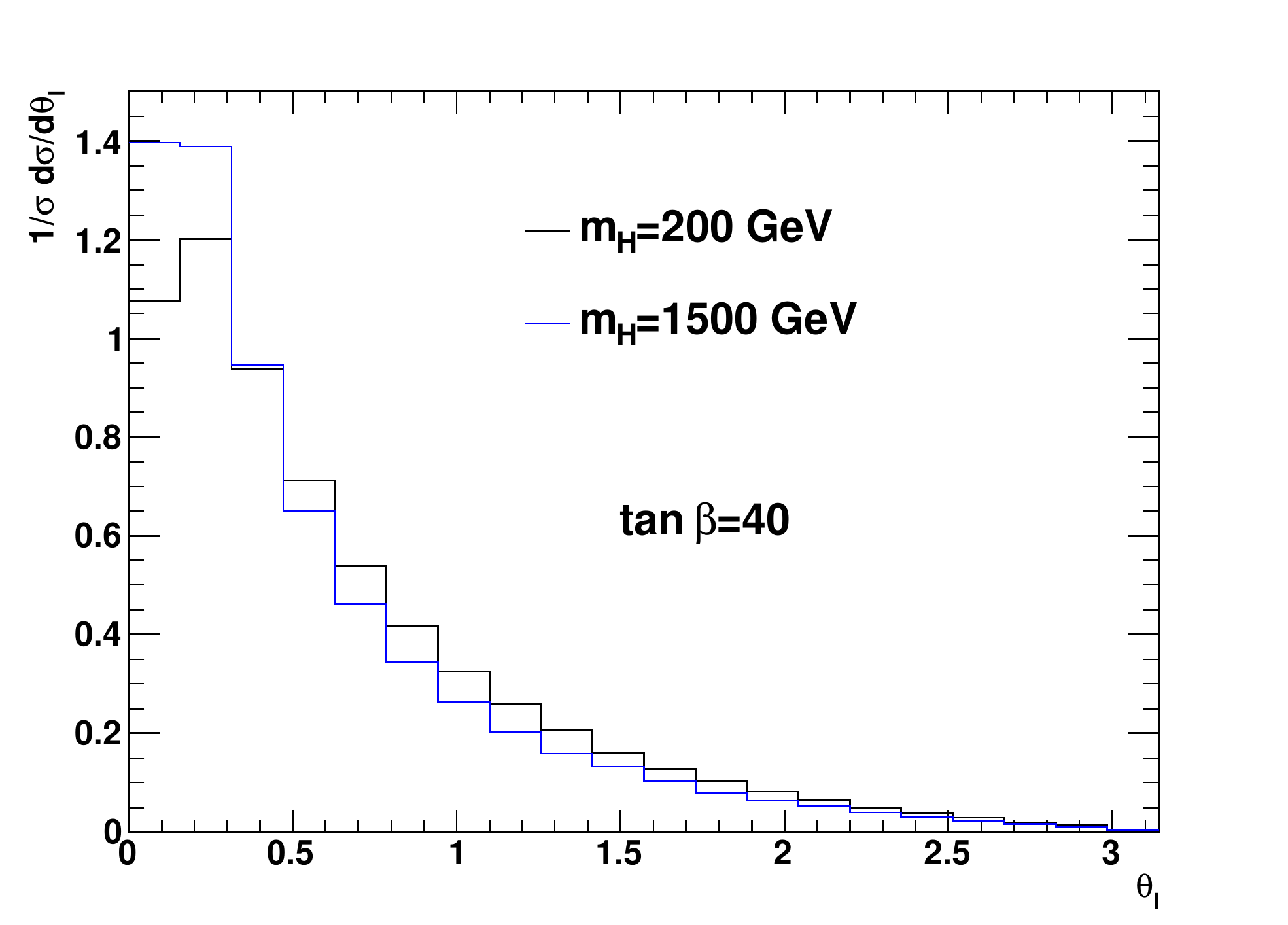}}
\caption{Polar angle $\theta_l$ between the top quark and its decay lepton in $H^-t$ production, obtained using MC@NLO~\cite{Weydert:2009vr}.}
\label{thetaplot}
\end{center}
\end{figure}

Results for the polar angle are shown in figure~\ref{thetaplot}. This is strongly peaked near $\theta_l=0$, again due to the boost from the top quark rest frame. Polarisation information also modifies the overall shape in this case, as can be encapsulated by the polar asymmetry parameter
\begin{equation}
      A_{\theta}=\frac{\sigma(\theta_l<\pi/4)-\sigma(\theta_l>\pi/4)}{\sigma(\theta_l<\pi/4)+\sigma(\theta_l>\pi/4)},
\end{equation}
whose behaviour is shown in figure~\ref{aplot}(b) for different parameter space values. We see in particular that the polar asymmetry is able to distinguish different charged Higgs boson masses at high $\tan\beta$, thus provides useful complementary information when combined with the azimuthal asymmetry.\\

\begin{figure}
\begin{center}
\scalebox{0.35}{\includegraphics{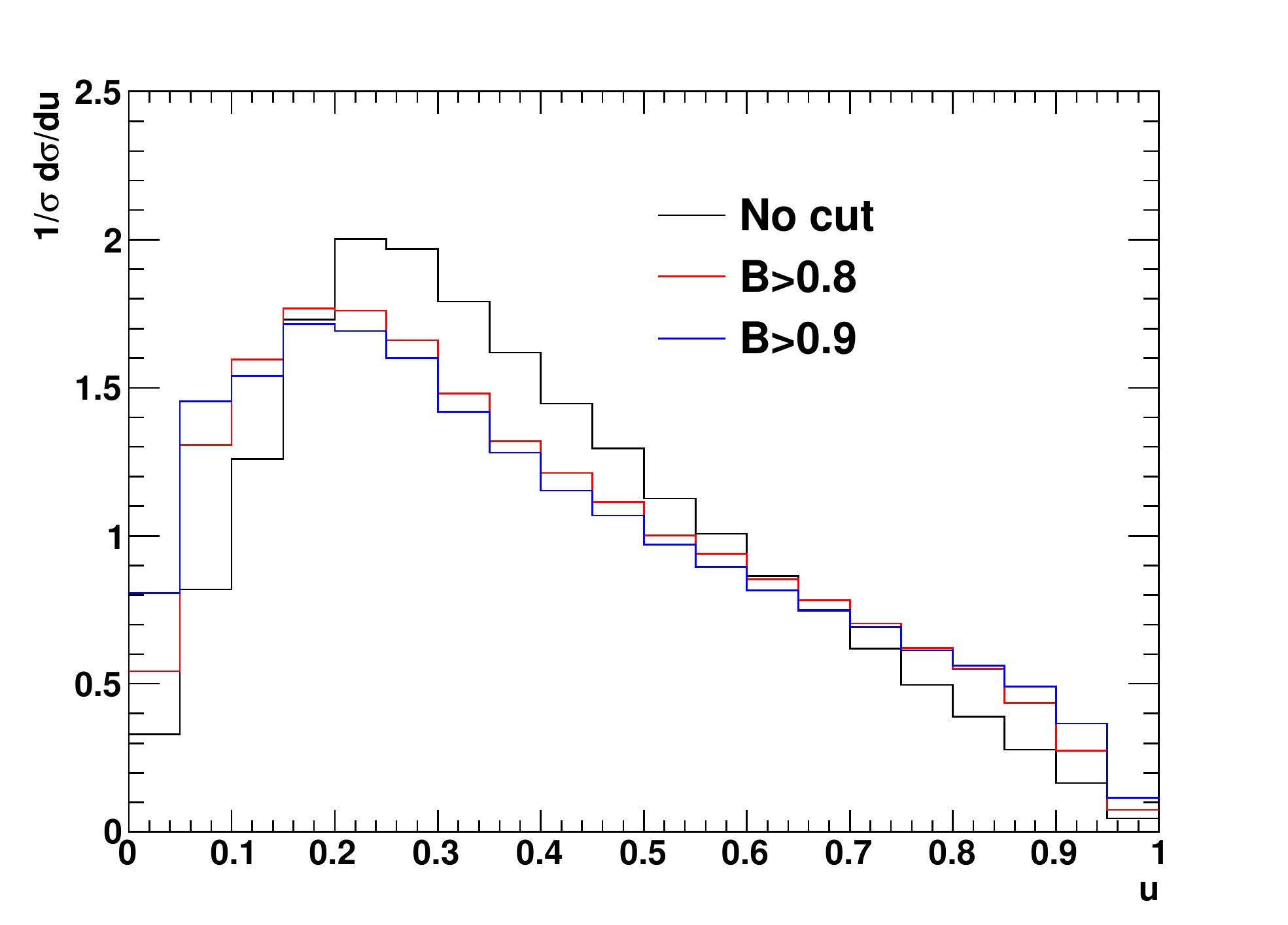}}
\scalebox{0.35}{\includegraphics{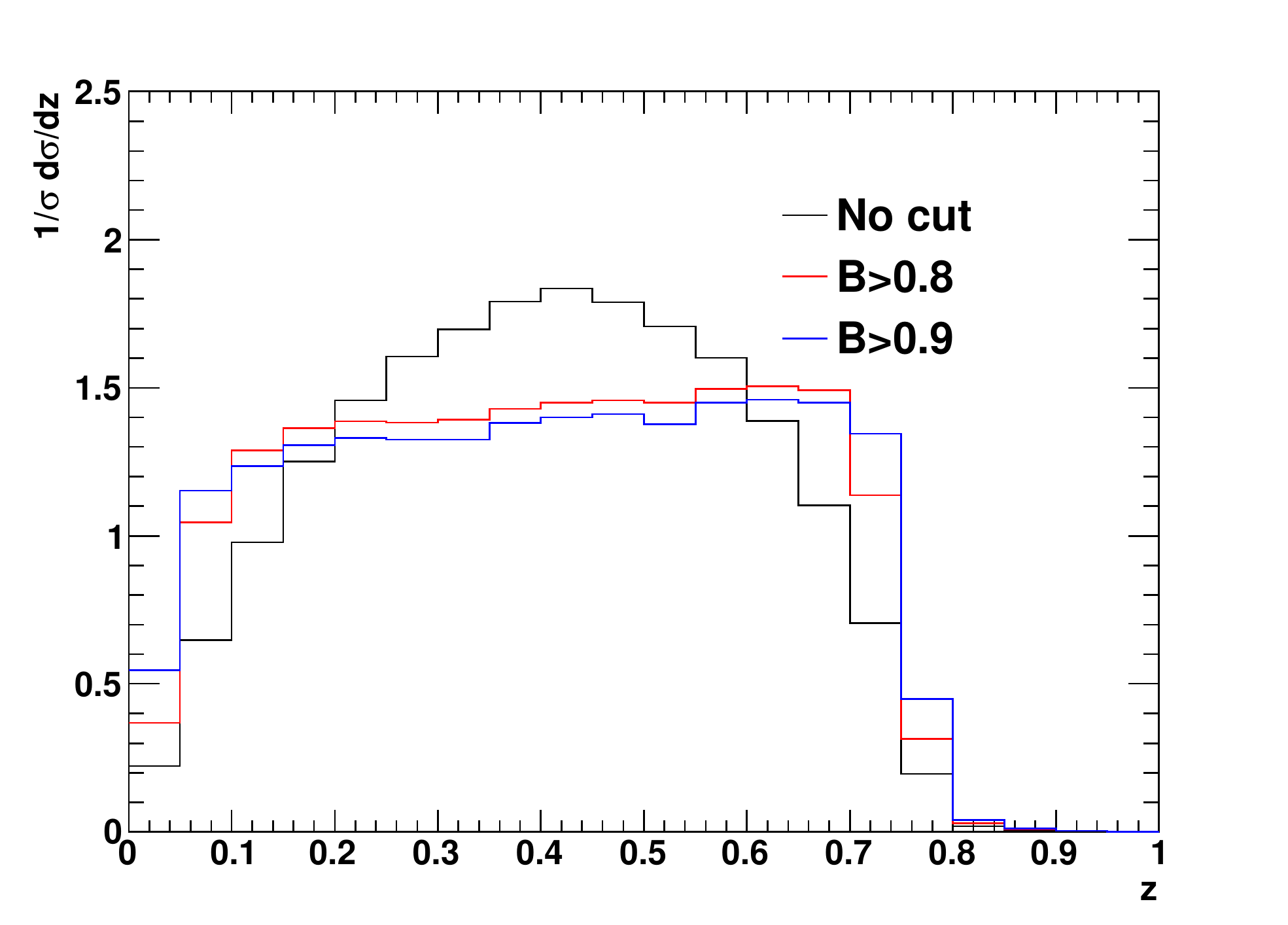}}
\caption{The $z$ and $u$ parameters in $H^-t$ production, obtained using MC@NLO~\cite{Weydert:2009vr}, and for different cuts on the boost parameter.}
\label{zuplots}
\end{center}
\end{figure}

\begin{figure}
\begin{center}
\scalebox{0.35}{\includegraphics{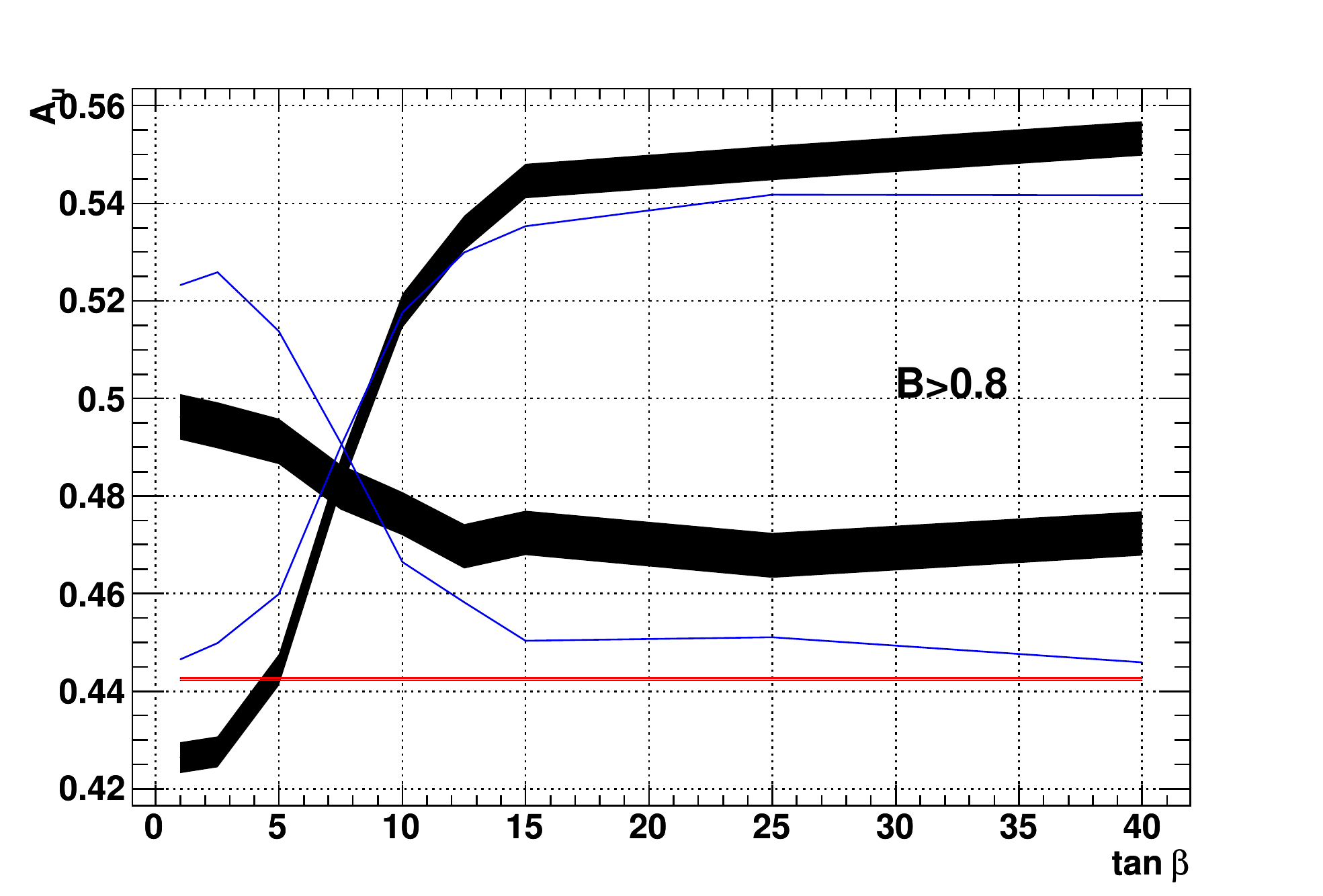}}
\scalebox{0.35}{\includegraphics{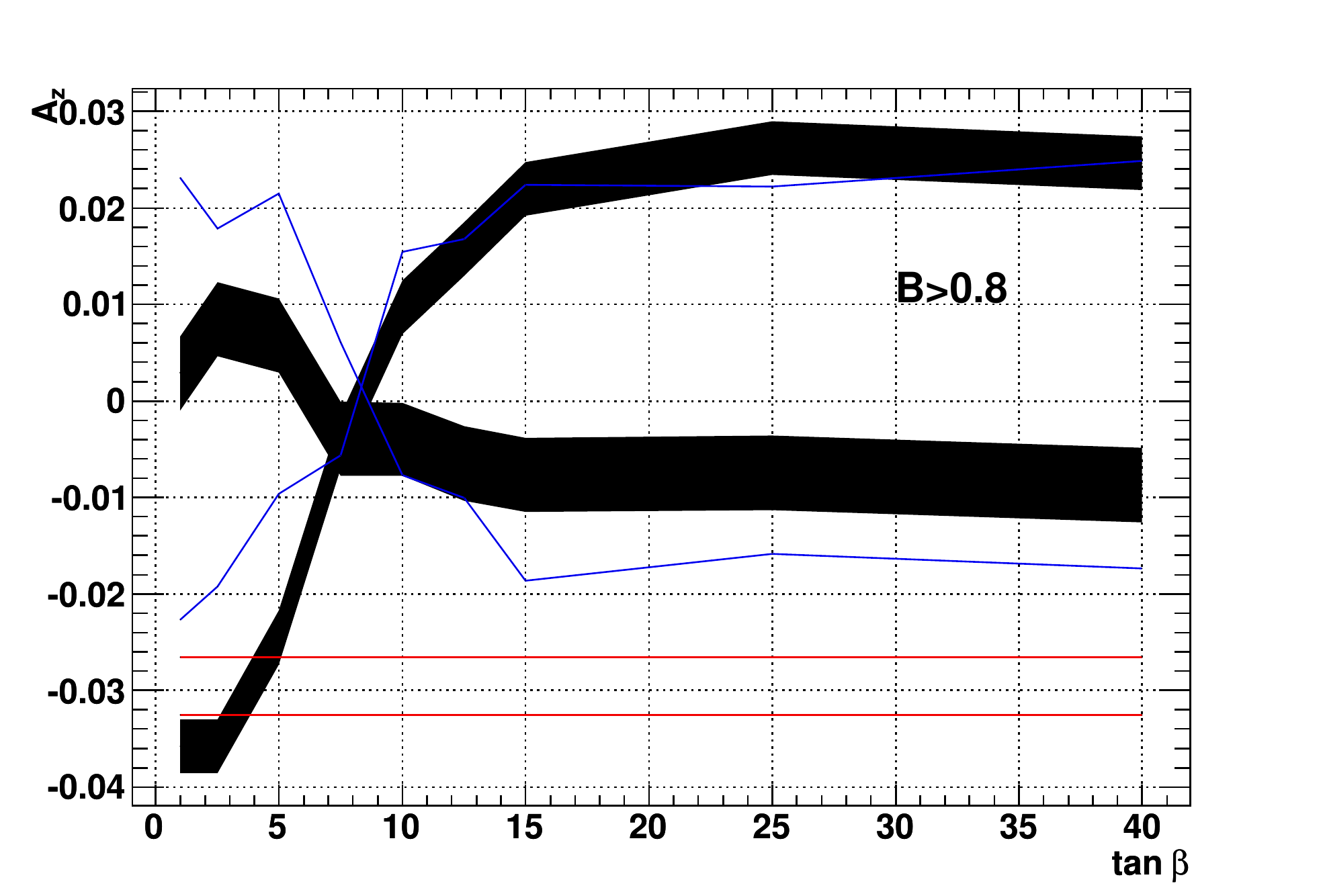}}
\caption{Energy asymmetry parameters, at both LO (blue) and MC@NLO (black) levels. Upper and lower curves correspond to charged Higgs boson masses of 1500 and 200 GeV. Also shown are the results obtained for SM $Wt$ production (flat bands).}
\label{azuplots}
\end{center}
\end{figure}

Further information can be gained by considering observables based on ratios of energies of the top quark and its decay products, which were shown in~\cite{Shelton:2008nq} to carry polarisation information:
\begin{equation}
      z=\frac{E_b}{E_t},\quad u=\frac{E_l}{E_l+E_b},
\label{zudef}
\end{equation}
Here $E_t$, $E_b$ and $E_l$ are the energies of the top quark, and its decay $b$ quark and lepton. The observables of eq.~(\ref{zudef}) were defined for boosted top quarks, such that it is easier to construct them. Here we characterise the boost of the top quark using the parameter
    \begin{equation}
    B=\frac{|\vec{p}_{\rm top}|}{E_t},
    \end{equation}
i.e. the ratio of three-momentum and energy. Upon imposing an increasing cut on the $B$ parameter, the shape of the distributions of the $z$ and $u$ parameters converge to a shape which carries polarisation information, as shown in figure~\ref{zuplots}.
Again there are crossing points for curves obtained at different values of $\tan\beta$ and $m_{H^-t}$. Analogously to the angular distributions considered earlier, one may define asymmetry parameters (see~\cite{Godbole:2011vw} for more details)
\begin{equation}
A_u=\frac{\sigma(u>0.215)-\sigma(u<0.215)}{\sigma(u>0.215)+\sigma(u<0.215)};
\quad
A_z=\frac{\sigma(0.1\leq z\leq 0.4)-\sigma(0.4\leq z\leq 0.7)}
{\sigma(0.1\leq z\leq 0.4)+\sigma(0.4\leq z\leq 0.7)}.
\label{Azudef}
\end{equation}
The behaviour of these asymmetries at different points in parameter space is shown in figure~\ref{azuplots}. 
One sees more sensitivity to additional radiation than for angular observables, although the results are still reasonably robust. Furthermore, the energy asymmetries provide complementary information to the angular observables: they are sensitive to new physics corrections in both the production and decay of the top quark, rather than just the production stage. A recent paper analysed the $z$ and $u$ parameters in a different new physics context, and also concluded that these observables remain useful even when detector effects and reconstruction ambiguities are accounted for~\cite{Papaefstathiou:2011kd}. 

To summarise, polarisation observables are an efficient way of probing the parameter space of new physics models. We have here focused on the case of $H^-t$ production. As stressed in~\cite{Godbole:2011vw}, similar methods can also be used in a purely SM context e.g. enhancing a signal of $Wt$ production against the top pair background.

\end{document}